\PassOptionsToPackage{table,x11names}{xcolor}
\documentclass{svjour3}

\usepackage{natbib}%
\usepackage{graphicx}%
\usepackage{multirow}%
\usepackage{amsmath,amssymb,amsfonts}%
\usepackage{mathrsfs}%
\usepackage[title]{appendix}%
\usepackage{xcolor}%
\usepackage{textcomp}%
\usepackage{manyfoot}%
\usepackage{booktabs}%
\usepackage{algorithm}%
\usepackage{algorithmicx}%
\usepackage{algpseudocode}%
\usepackage{listings}%
\usepackage[normalem]{ulem}

\algnewcommand\algorithmicforeach{\textbf{for each}}
\algdef{S}[FOR]{ForEach}[1]{\algorithmicforeach\ #1\ \algorithmicdo}

\usepackage[cleanlook,british]{isodate}
\usepackage[coloredits,editnote,unblinded,roman,todo,draftkeys,nodraftbg]{mycommon}

\usepackage{graphicx}

\usepackage{csquotes}
\usepackage{booktabs,tabularx,multirow,multicol}
\usepackage[inline]{enumitem}
\usepackage{subcaption}

\usepackage{tikz}
\usetikzlibrary{shapes,positioning,arrows.meta,decorations.pathreplacing}

\usepackage{listings}

\lstdefinelanguage{Kotlin}{
  comment=[l]{//},
  emph={filter, first, firstOrNull, forEach, lazy, map, mapNotNull, println},
  keywords={!in, !is, abstract, actual, annotation, as, as?, break, by, catch, class, companion, const, constructor, continue, crossinline, data, delegate, do, dynamic, else, enum, expect, external, false, field, file, final, finally, for, fun, get, if, import, in, infix, init, inline, inner, interface, internal, is, lateinit, noinline, null, object, open, operator, out, override, package, param, private, property, protected, public, receiveris, reified, return, return@, sealed, set, setparam, super, suspend, tailrec, this, throw, true, try, typealias, typeof, val, var, vararg, when, where, while},
  morecomment=[s]{/*}{*/},
  morestring=[b]",
  morestring=[s]{"""*}{*"""},
  ndkeywords={@Deprecated, @JvmField, @JvmName, @JvmOverloads, @JvmStatic, @JvmSynthetic, Array, Byte, Double, Float, Int, Integer, Iterable, Long, Runnable, Short, String},
  sensitive=true,
  escapeinside={/*@}{@*/}
}
\lstdefinestyle{default}{
  basicstyle=\ttfamily,
  breaklines=true,
  prebreak=\raisebox{0ex}[0ex][0ex]{\ensuremath{\hookleftarrow}},
  numbers=left,
  numberstyle=\scriptsize,
  xleftmargin=0.4cm
}

\lstnewenvironment{kotlin}
  {\lstset{language=Kotlin,style=default}}
  {}

\lstnewenvironment{java}
  {\lstset{language=Java,style=default,morekeywords={var},escapeinside={/*@}{@*/}}}
  {}

\newcommand{\codebox}[1]{\fcolorbox{black!50}{gray!5}{#1}}
\newcommand{\codeboxreq}[1]{\fcolorbox{red!40}{red!5}{#1}}

\usepackage{hyperref}
\hypersetup{hidelinks}
\makeatletter
\def\cl@chapter{\@elt {theorem}}
\makeatother
\usepackage[nameinlink,capitalize]{cleveref}

\begin{document}

\title{How Do Developers Use Type Inference: An Exploratory Study in Kotlin}

\author{Samuel W. Flint \and Ali M. Keshk \and\\ Robert Dyer \and Hamid Bagheri}
\institute{
  Corresponding author: Samuel W. Flint \at
  University of Nebraska--Lincoln\\
  \email{swflint@huskers.unl.edu}
  \and
  Ali M. Keshk \at
  University of Nebraska--Lincoln\\
  \email{akeshk2@huskers.unl.edu}
  \and
  Robert Dyer \at
  University of Nebraska--Lincoln\\
  \email{rdyer@unl.edu}
  \and
  Hamid Bagheri \at
  University of Nebraska--Lincoln\\
  \email{bagheri@unl.edu}
}

\maketitle

\abstract{Statically typed languages offer numerous benefits to developers, such as improved code quality and reduced runtime errors, but they also require the overhead of manual type annotations.
To mitigate this burden, language designers have started incorporating support for type inference, where the compiler infers the type of a variable based on its declaration/usage context.
As a result, type annotations are optional in certain contexts, and developers are empowered to use type inference in these situations.
However, the usage patterns of type annotations in languages that support type inference are unclear.
These patterns can help provide evidence for further research in program comprehension, in language design, and for education.
We conduct a large-scale empirical study using Boa, a tool for mining software repositories, to investigate when and where developers use type inference in 498,963 Kotlin projects.
We choose Kotlin because it is the default language for Android development, one of the largest software marketplaces.
Additionally, Kotlin has supported declaration-site optional type annotations from its initial release.
Our findings reveal that type inference is frequently employed for local variables and variables initialized with method calls declared outside the file are more likely to use type inference.
These results have significant implications for language designers, providing valuable insight into where to allow type inference and how to optimize type inference algorithms for maximum efficiency, ultimately improving the development experience for developers.}

\section{Introduction}
\label{sec:introduction}

Static typing aids developers in a number of ways: verification of type safety~\citep{pacak20:_system_approac_deriv_increm_check}, detection of errors~\citep{prechelt98:_contr_exper_asses_benef_proced,kleinschmager12:_do}, and as in-code documentation~\citep{lubin21:_how,hanenberg14:_empir_study_impac_static_typin_softw_maint}.
However, static typing can present a maintenance burden: requiring that types are annotated potentially increases cognitive load~\citep{chalin07:_non_refer_defaul_java,ore18:_asses_type_annot_burden}.
Additionally, static type systems may slow developers down when completing some tasks~\citep{hanenberg10:_exper_static_dynam_type_system,okon16:_can,stuchlik11:_static_vs}, though evidence is mixed~\citep{mayer12:_empir_study_influen_static_type,prechelt98:_contr_exper_asses_benef_proced,kleinschmager_etal12:_do_static_type_system_improv}.
To simplify code or reduce clutter~\citep{basic-modern-cpp}, many statically typed languages (such as C++, Java, C\#, Scala, Swift, and Kotlin) have adopted type inference, which allows type annotations to be dropped in some circumstances.

Previous work has examined type systems and their effects~\citep{mayer12:_empir_study_influen_static_type} and how developers use and conceptualize types in static type systems~\citep{lubin21:_how}.
However, as of writing, we are unaware of any work that studies in-depth how developers use type inference (though much work has been done to improve its performance~\citep{hellendorn18:_deep,abraham06:_type,pierce00:_local}).
Understanding where developers use type inference can help researchers better target type inference for future, human-subjects studies.
In particular, understanding the context of addition, or removal, of type annotations (i.e., changing from or to using type inference) can help to build better models of developer cognition.
  Additionally, knowing the sorts of expressions, and locality of calls, commonly used with inferred types may be helpful to the developers and implementers of type inference algorithms, providing information about what to cache or how better to store the relevant information.

To achieve these goals, in this paper, we perform an exploratory study on the use of the Kotlin language.
Kotlin is a top-ten statically typed programming language~\citep{tiobe,pypl,githuboctoverse} and is the default programming language for Android development (the most popular mobile platform in the world).
The Kotlin language has supported type inference in various forms since its initial release in 2011.
Additionally, because of Kotlin's strong interoperability with Java, we are able to study how that interoperability might affect the use of type inference.

This paper offers two main contributions.
First, while prior work considered how optional typing was used in dynamic languages~\citep{souza14:_how,di22:_evolut_type_annot_python}, this is the first study to investigate the usage of type inference in statically typed languages.
Second, we examine the influence of different development contexts on the use of type inference.

To study the use of type inference in Kotlin, we utilize the Boa infrastructure~\citep{dyer13:_boa} to mine a large GitHub dataset consisting of almost 500k open-source Kotlin projects.
First, we investigate how often developers utilize type inference (where it is available to them).
Since type inference often involves assignment expressions, we analyze the right-hand sides of assignments to see what kinds of expressions are most common.
Next, we see if there are differences in the choice to use type inference in other contexts, such as mutable (\lstinline[language=Kotlin]|var|) versus immutable (\lstinline[language=Kotlin]|val|) variables, testing versus non-testing code, and when projects also contain Java source files alongside the Kotlin source files.
Finally, we consider the survivability of inference use: when type inference is (not) used, what is the probability of that changing over time?
This particular analysis can help to understand whether or not type inference is used for the purposes of documentation.

The main findings of our study are as follows:

\begin{itemize}
  \item Kotlin developers use type inference the most for local variables, fields, and method return types, implying this is a useful feature for future programming languages;
  \item in Kotlin, most local variables utilize type inference while most fields do not, implying locality of the declaration plays a role;
  \item initializers of type-inferred variables are typically calls to non-file-local methods, implying that type inference engines need to have global type knowledge; and
  \item type annotation presence (or absence) on variables tends to avoid changing over time, implying inference tools can rely on caching more often.
\end{itemize}

\noindent Given these findings, this paper makes the following contributions:

\begin{itemize}
    \item A first-of-its-kind characterization of the usage of declaration-level type inference by developers in Kotlin;
    \item a description of common initializer expressions used for type-inferred variables; and
    \item a survival analysis characterizing change patterns of type annotations in code.
\end{itemize}

In \Cref{sec:background} we provide background for type inference and how it is used in Kotlin.
We then describe our research questions in \Cref{sec:research-questions} and our methods in \Cref{sec:methods}.
Next, we provide our results in \Cref{sec:results} and consider the limitations in \Cref{sec:threats-validity}.
We discuss the implications of the results in \Cref{sec:discussion} and consider the relation to prior works in \Cref{sec:prior-work}.
Finally, we conclude in \Cref{sec:conclusion}.

\section{Background}
\label{sec:background}

In this section, we provide some brief background on the Kotlin language, discuss type inference generally, and discuss where type inference can be used in Kotlin.

\subsection{Kotlin Language}
\label{sec:kotlin-background}

As readers may not be familiar with Kotlin, we provide some brief background on some of its language features.
Kotlin targets the Java VM with full interoperability with existing Java libraries.
It is a statically typed language but also supports type inference in several locations (see \cref{sec:type-infer-kotl}).
One of the goals of the language design was to make it easier for developers to write code that is more concise than typically written in Java, by providing specialized syntax and various features like data classes, destructuring, and support for building domain-specific languages (DSLs).

\begin{figure}
\begin{kotlin}
// global variables
var g/*@\codebox{: \textbf{Int}}@*/ = 7 /*@\label{kt:top-level}@*/

// class/object fields (properties are implicit, and ignored no matter their inference status)
class C {
  var x/*@\codebox{: \textbf{Int}}@*/ = 5 /*@\label{kt:class-field}@*/
}

// return types of single-expression functions
fun square(x/*@\codeboxreq{: \textbf{Int}}@*/)/*@\codebox{: \textbf{Int}}@*/ = x * x /*@\label{kt:single-exp-ret}@*/

fun hypotenuse(x/*@\codeboxreq{: \textbf{Int}}@*/, y/*@\codeboxreq{: \textbf{Int}}@*/)/*@\codeboxreq{: \textbf{Double}}@*/ { /*@\label{kt:block-ret1}@*/
  // local variable declarations
  val x2/*@\codebox{: \textbf{Double}}@*/ = square(x).toDouble() /*@\label{kt:mtd-defn1}@*/
  val y2/*@\codebox{: \textbf{Double}}@*/ = square(y).toDouble() /*@\label{kt:mtd-defn2}@*/
  return Math.sqrt(x2 + y2)
}

fun makelist(x/*@\codeboxreq{: \textbf{Int}}@*/)/*@\codeboxreq{: \textbf{MutableList}<\textbf{Int}>}@*/ { /*@\label{kt:block-ret2}@*/
  var lst/*@\codebox{: \textbf{MutableList}<\textbf{Int}>}@*/ = mutableListOf<Int>() /*@\label{kt:mtd-defn3}@*/
  // loop iterator variables
  for (i/*@\codebox{: \textbf{Int}}@*/ in 1..x) /*@\label{kt:loop-iterator}@*/
    lst.add(i)
  return lst
}

// lambda argument lists
fun factN(n/*@\codeboxreq{: \textbf{Int}}@*/) = (makelist(n) as /*@\codeboxreq{\textbf{List<Int>}}@*/).fold(1) /*@\label{kt:cast}@*/ 
             { acc/*@\codebox{: \textbf{Int}}@*/, i/*@\codebox{: \textbf{Int}}@*/ -> acc * i } /*@\label{kt:lambda-args}@*/

fun main() {
  // Implicit Destructuring Declarations
  for ((a/*@\codebox{: \textbf{Int}}@*/, b/*@\codebox{: \textbf{Int}}@*/) in makelist(g).map({ x/*@\codebox{: \textbf{Int}}@*/ -> Pair(x, factN(x)) })) { /*@\label{kt:implicit-destructure}@*/
    println("$a,$b")
  }
}
\end{kotlin}
\caption{Example type inference locations in Kotlin. Developers can omit code in \codebox{gray} and the compiler will try to infer the type.
  Type annotations in \codeboxreq{red} do not support inference and must be supplied by the developer.}
\label{fig:type-inference-examples}
\end{figure}

A full discussion of Kotlin's language is out of the scope of this work, but here we mention a few relevant features.
Kotlin allows developers to control variable mutability by choosing between \emph{immutable} variables that may only be assigned to once (\lstinline[language=kotlin,style=default]|val|, \eg, \Cref{fig:type-inference-examples} line~\ref{kt:mtd-defn1}) and \emph{mutable} (\lstinline[language=kotlin,style=default]|var|, \eg, line~\ref{kt:top-level}) keywords when declaring variables.
There is also support for \enquote{single-expression functions} (line~\ref{kt:single-exp-ret}), denoted using an equal sign rather than curly braces. Such functions implicitly return the value computed by the expression on the right-hand side of the definition.

\subsection{Type Inference}
\label{sec:bg-type-inference}

Type inference is often used in the context of statically typed languages, though some implementations of dynamically typed languages use it for optimization.
Static type systems can be relatively simple (\eg, nominal type systems) or more complex (\eg, structural type systems).
In either case, static typing generally presents several benefits to developers~\citep{mayer12:_empir_study_influen_static_type,stuchlik11:_static_vs}, such as a potential documentary effect~\citep{spiza14:_type_apis}, memory support~\citep{lubin21:_how}, or as a useful prototyping and domain modeling tool~\citep{lubin21:_how}.
Static type systems can also be used to ensure code correctness in various ways~\citep{colazzo04:_types_path_correc_xml_queries} or support memory management~\citep{montenegro08:_type_system_safe_memor_manag} (e.g., Rust's ownership system~\citep{klabnik22:_rust_progr_languag}).

Such benefits come with a trade-off, requiring users to manually annotate the type of every variable in the system.
Type inference provides a way to help balance the benefits and the trade-offs, by allowing developers to omit specifying type annotations in certain places.
The compiler then statically determines the type using constraint-solving techniques.
Next, we describe how and where type inference is supported in Kotlin.

\subsection{Type Inference in Kotlin}
\label{sec:type-infer-kotl}

Because the focus of this study is the use of type inference, we consider specifically the locations in Kotlin code where type inference is allowed.
The official Kotlin specification~\citep{akhin20:_type_infer} describes these locations as follows (examples are shown in \Cref{fig:type-inference-examples} in gray boxes):

\begin{itemize}
\item global variables (line~\ref{kt:top-level}),
\item class/object fields (line~\ref{kt:class-field}),
\item return types of single-expression functions (line~\ref{kt:single-exp-ret}),
\item local variable declarations (lines~\ref{kt:mtd-defn1}--\ref{kt:mtd-defn2}, \ref{kt:mtd-defn3}),
\item loop iterator variables (lines~\ref{kt:loop-iterator} and \ref{kt:implicit-destructure}), and
\item lambda arguments lists (line~\ref{kt:lambda-args}).
\end{itemize}

Certain locations do not support type inference and developers are required to provide a type annotation.
Many of these locations also implicitly mark the variables as immutable (\ie, implicitly declared with the \lstinline[language=kotlin,style=default]|val| keyword).
Some examples of locations where type inference is not supported (examples are shown in \Cref{fig:type-inference-examples} in red boxes):

\begin{itemize}
    \item method/function parameters (lines~\ref{kt:single-exp-ret}, \ref{kt:block-ret1}, \ref{kt:block-ret2})
    \item casts (line~\ref{kt:cast})
    \item return type of functions with block bodies (lines~\ref{kt:block-ret1}, \ref{kt:block-ret2})--note that the type can be omitted, but then it is always assumed to be of type \texttt{Unit}
\end{itemize}

Finally, in declarations where type inference is otherwise permitted, to use type inference an initializer expression \emph{must} be present.
If an initializer expression is not present, then a type annotation must be provided.
So while immutable declarations do allow deferred assignment, if the assignment is deferred, the declaration site must contain a type annotation.

\section{Methodology}
\label{sec:methods}

In this section, we describe the dataset used (\Cref{sec:dataset}), filtering and data preparation (\Cref{sec:data-preparation}), the measurements used (\Cref{sec:measurements}), and our analysis methods (\Cref{sec:analysis-methods}).

\subsection{Research Questions}
\label{sec:research-questions}

As type inference is becoming increasingly popular in modern programming languages, understanding where and how often developers utilize it can help provide insights into future language designs.
To that end, we study the Kotlin language, the default programming language for Android, the most popular mobile operating system.

\begin{rqs}
\item\label{rq:usage} \textbf{How often do Kotlin developers use type inference?}
  Kotlin supports type inference in specific locations, but developers are not required to use it.
  Understanding the frequency of use of type inference may help to understand where developers prefer manually adding type annotations.
\end{rqs}

Based on the results observed when answering this question, and because different code contexts present different challenges to developers, we then look deeper to see if certain contexts affect the use of type inference.

\begin{rqs}
\item\label{rq:context} \textbf{Is type inference used differently in different contexts?}
  Where type inference is used, does that use differ depending on the development context?
  Here we consider four sub-questions, corresponding to four different contexts we observed:

  \begin{subrqs}
  \item\label{rq:rhs} \textbf{Is type inference used more frequently with certain kinds of initializer expressions?}
    Here we consider local variables, and the types of expressions on the right-hand side (\eg, literal, variable access, method call) to see what expression kinds are used more frequently.
      
  \item\label{rq:mutability} \textbf{Is the use of type inference different for immutable variables (declared with \texttt{val}) or mutable variables (declared with \texttt{var})?}
    Is there a relationship between mutability and the usage of type inference?
      
  \item\label{rq:tests} \textbf{Is the usage of type inference different in testing classes?}
    As the goals of writing tests are different from writing non-testing code, the usage of type inference may also be different.
      
  \item\label{rq:mixed-projects} \textbf{Is the usage of type inference different in Kotlin projects containing both Kotlin and Java source files?}
    Since Kotlin projects can also include Java source files, and Java supports type inference in fewer locations, the presence of Java source files may influence the behavior of Kotlin developers.
  \end{subrqs}
\end{rqs}

Finally, based on our personal experience, we have rarely seen code with a type annotation remove that type over time.
What is less clear, however, is if variables utilizing type inference (that is, which are originally declared without a type annotation) eventually add a type annotation.
Reasons to do so might be for documentation or to place additional constraints on the inference algorithm to help it infer the correct types elsewhere.

\begin{rqs}
  \item\label{rq:over-time} \textbf{Does the use of type inference change over time by adding or removing type annotations?}
\end{rqs}

\subsection{Dataset}
\label{sec:dataset}

We use the \enquote{2021 Aug/Kotlin} Boa~\citep{dyer13:_boa,boa-website} dataset (containing 499,645 projects), which is composed of nearly all of the open-source Kotlin projects publicly available from GitHub at the time the dataset was cloned in the summer of 2021.
Boa builds datasets by using the GitHub API to find project metadata and then cloning the Git repositories for those projects. It then converts that data into a custom AST format for easier querying.
This dataset is selected because of its usability and broadly representative nature.
Within the dataset, we find small (even toy) projects as well as large, well-known Kotlin projects, such as \texttt{leakcanary},\footnote{\url{https://github.com/square/leakcanary}} Shadowsocks for Android,\footnote{\url{https://github.com/shadowsocks/shadowsocks-android}} and the \texttt{anko} library for Android development.\footnote{\url{https://github.com/Kotlin/anko/}}
We include projects of all sizes to see if there is a relationship between project size and the usage of type inference.

\begin{table}[htbp]
  \centering
  \caption{Analyzed repositories and files (file-level deduplication not accounted for).}
  \label{tab:analyzed-repos}
  \small
\begin{tabular}{lrr}
  \toprule
   & \textbf{Raw Dataset} & \textbf{Filtered} \\
  \midrule
  \textbf{Projects} & 499,645 & 498,963 \\
  \textbf{Source Files (HEAD)} & 9,604,478 & 9,604,206 \\
  \textbf{Snapshots} & 37,866,871 & 37,865,222 \\
  \bottomrule
  \end{tabular}
    \vspace{-1em}
\end{table}

The number of raw projects in the dataset is shown in \Cref{tab:analyzed-repos}, and the number of commits, files, and statements are shown in \Cref{tab:repo-size-stats}.
Median Kotlin projects tend to be young (3 commits) and small (9 files).

\begin{table*}[htbp]
  \centering
  \caption{Repository size statistics.}
  \label{tab:repo-size-stats}
  \small
  \renewcommand{\tabcolsep}{0.5em}
\begin{tabular}{lrrrrrrr}
  \toprule
   & \textbf{mean} & \textbf{std} & \textbf{min} & \textbf{25\%} & \textbf{50\%} & \textbf{75\%} & \textbf{max} \\
  \midrule
  \textbf{Main Branch Commits} & 14.80 & 114.11 & 1 & 1 & 3 & 10 & 14,723 \\
  \textbf{Number of Committers} & 2.02 & 34.59 & 0 & 1 & 1 & 2 & 7,724 \\
  \textbf{Number of Files} & 19.25 & 77.15 & 1 & 4 & 9 & 20 & 36,021 \\
  \textbf{Number of Files (Deduped)} & 8.53 & 64.57 & 1 & 2 & 2 & 5 & 11,998 \\
  \textbf{Number of Statements} & 406.00 & 2,484.82 & 1 & 46 & 125 & 330 & 1,101,330 \\
  \textbf{Number of Stars} & 3.33 & 135.21 & 0 & 0 & 0 & 0 & 38,979 \\
  \bottomrule
  \end{tabular}
\end{table*}

\subsection{Filtering and Data Preparation}
\label{sec:data-preparation}

To ensure that detected uses of type inference are relevant, we filter some projects from the dataset.
We filtered projects which had no parseable Kotlin files or for which the only Kotlin file was \texttt{build.gradle.kts}.

After filtering, we prepared the data using the Boa Study Template~\citep{dyer22:_boa_study_templ}.
Additionally, for some research questions (\ref*{rq:rhs} and \ref*{rq:over-time}), we removed duplicates at the file level, as suggested by \citet{lopes17:_dejav}.
For this, we collected the AST hash of each file in the HEAD snapshot of each project and selected for analysis only one file-project pair for each hash.

\subsection{Measurements}
\label{sec:measurements}

We collect data about five features:
\begin{enumline}
\item declaration site
\item usage of type inference
\item initializer expression type
\item initializer method call locality
\item time from creation to change in inference status.
\end{enumline}
Additionally, for \ref*{rq:usage} and \ref*{rq:context} we consider each repository's HEAD commit snapshot.
For \ref*{rq:over-time}, we consider changes over time.

\paragraph{RQ1: Declaration Site Detection}
As type inference is not allowed in all variable declaration locations, we must be careful to only mine those where it is available.
This is done through selectively visiting explicit variable declarations, based on the locations which are available as described in \Cref{sec:type-infer-kotl}.

\paragraph{RQ1: Detecting Type Inference}
We detect whether or not a given declaration site uses a type annotation by analyzing the Boa AST.
In Boa's AST, all variable declarations are represented as a \texttt{Variable} object, which contains a \texttt{type} field.  If this field is defined, the variable has a type annotation and is thus not inferred.
If instead the \texttt{type} field is undefined, the variable declaration uses type inference.

\paragraph{RQ2(a): Detecting Initializer Expression Types}
We also collect the kind of initializer expressions (when present).
That is, what sort of expression is present, not the type of that expression.
This information is partially available in the Boa ASTs, however determining if a call is to a method or a constructor is not straightforward, as the syntax for both is identical in Kotlin, and type information (which Boa does not provide) is required to resolve this.

To resolve the difference between calls to methods and constructors, we use a two-pass approach.
On the first pass, we collect the names of all project-local methods and classes into two separate sets.
In the second pass, we use the following heuristic: if a method call name is part of the list of project-local methods or default-imported methods, we consider it to be a method call.
If this first condition fails, we check to see if this is a default-imported, file-imported, or project-local class.
If so, we consider it to be a constructor call.
Finally, if the prior two conditions failed, we use a name-starts-with-capital-letter heuristic: if the name starts with a capital letter, we assume it to be a constructor call, otherwise, we assume it to be a method call.

\paragraph{RQ2(b): Detecting Method Call Locality}
Similar to the above issue with detecting method calls or constructor calls in Kotlin, we also detect whether calls to methods are \textit{file-local} (the method is declared in the same file) using a simple heuristic.
While to do so perfectly would require complete type information (and thus be infeasible using the Boa infrastructure), our heuristic is complete with respect to non-local calls.
For each file, we collect the names of defined methods and as we come to variable declarations with method calls in their initializers, we check to see if the name is in this list.
If so, we consider it a \enquote{possibly file-local call}, if not, we consider it as a \enquote{non-file-local call}.

\paragraph{RQ2(c): Detecting Tests}
To determine if a file is a test, we use two heuristics.
The first (and simplest) is a test of the file path: if the lowercased path of a source file contains either \texttt{test} or \texttt{testing}, or ends in \texttt{test.kt}, the file is considered a test.
This is a similar heuristic that has been applied previously in many other studies~(e.g., \cite{Vahabzadeh15,Veloso22,Nagy22}; \cite{Coro20,nakamaru20,keshk23}).
The second heuristic is somewhat more complicated: if a file imports one of a number of popular testing frameworks (in the testing or mocking categories on MVN Repository), it is considered a test.
Our list of testing frameworks also includes popular Java testing frameworks (as Kotlin projects may use Java libraries).
This is a similar approach that has also been used in previous studies (e.g., \cite{Peruma21,Nagy22}).
The frameworks included:\footnote{\url{https://github.com/unl-pal/kotlin-inference-study/blob/af15202888b4f1aac6c9d748137cf7c0e6720b1e/boa/queries/common/detect-tests.boa\#L5}}

\begin{itemize}
    \item Java/Kotlin Frameworks:
    \begin{itemize}
        \item org.junit.*
        \item org.scalatest.*
        \item org.testng.*
        \item org.springframework.test.*
        \item org.springframework.mock.*
        \item org.hamcrest.*
        \item org.scalacheck.*
        \item org.mockito.*
        \item org.easymock.*
        \item org.powermock.*
        \item com.github.tomakehurst.*
        \item org.easymock.*
        \item org.jmock.*
        \item org.jmockit.*
    \end{itemize}
    \item Kotlin Frameworks:
    \begin{itemize}
        \item kotlin.test.*
        \item io.mockk.*
        \item org.spekframework.*
        \item com.natpryce.hamkrest.*
        \item io.kotest.*
        \item io.strikt.*
        \item org.amshove.kluent.*
        \item com.winterbe.expekt.*
        \item assertk.*
    \end{itemize}
\end{itemize}

\paragraph{RQ2(d): Detecting Mixed Kotlin-Java Projects}
When investigating \ref*{rq:mixed-projects}, we must be able to determine if a project contains Java code.
To do this, we collect a list of all projects with at least one (parseable) Java source file in their HEAD snapshot and filter prior query results using this list.

\paragraph{RQ3: Changing Inference Status}
To investigate \ref*{rq:over-time}, we collect information about type inference usage over time.
In particular, we collect the lifetime of a declaration's type inference status, that is: how long did the initial declaration stay, and how did it change, if at all.
To do this, we process all commits reachable from \texttt{HEAD} in the order they were committed.  The full algorithm is shown in \Cref{alg:infer-status}.

\begin{algorithm}
  \scriptsize
  \caption{Algorithm for determining inference status over time.}
  \label{alg:infer-status}
    \begin{algorithmic}[1]
    \Procedure{ProcessRepository}{reachableCommits}
      \State allDeclarations $\gets$ \{\}\Comment{Map from file to declarations}
      \ForEach{commit $\in$ reachableCommits}
        \ForEach{file $\in$ commit}
          \State knownDeclarations $\gets$ allDeclarations[file]\Comment{List of statuses}
          \State timeDifference $\gets$ \textsc{time}(commit) - \textsc{createTime}(knownDeclarations, decl)
\State
          \ForEach{decl \(\in\) knownDeclarations $and$ $\not\in$ file}\Comment{Find removed declarations}
            \State knownDeclarations[decl] $\gets$ \{REMOVED, \textsc{inferred?}(decl), timeDifference\}
          \EndFor
\State
          \ForEach{decl $\in$ file $and$ $\not\in$ \textsc{ignored}(knownDeclarations, decl)}\
            \If{decl $\not\in$ knownDeclarations}\Comment{The declaration is \emph{new}}
              \State knownDeclarations[decl] $\gets$ \{ADDED, \textsc{inferred?}(decl), timeDifference\}
            \Else
              \Statex\Comment{The declaration's inference status changed}
              \If{\Call{initialInferred}{knownDeclarations, decl} $\neq$ \Call{inferred?}{decl}}
                \State knownDeclarations[decl] $\gets$ \{CHANGED, \textsc{inferred?}(decl), timeDifference\}
              \EndIf
            \EndIf
          \EndFor
\State
          \If{commit $is$ $last$ $commit$}\Comment{Mark last observation times}
            \ForEach{decl $\in$ knownDeclarations}
              \If{decl $\not\in$ \textsc{ignored}(knownDeclarations, decl)}
                \State knownDeclarations[decl] $\gets$ \{OBSERVATION\_ENDED, \textsc{inferred?}(decl), timeDifference\}
              \EndIf
            \EndFor
          \EndIf
        \EndFor
      \EndFor
    \EndProcedure
  \end{algorithmic}
\end{algorithm}

\begin{algorithm}
  \scriptsize
  \caption{Finding the initial inference status.}
  \label{alg:initialInferred}
  \begin{algorithmic}[1]
    \Procedure{initialInferred}{$knownDeclarations$, $decl$}
      \If{decl $\in$ knownDeclarations}
        \ForEach{observation \(\in\) knownDeclarations[decl]}
          \If{observation.kind $is$ ADDED}
            \State \Return observation.inferred?
          \EndIf
        \EndFor
      \EndIf
      \State \Return \textsc{undefined}
    \EndProcedure
  \end{algorithmic}
\end{algorithm}

\begin{algorithm}
  \scriptsize
  \caption{Find time of first observation, if any.}
  \label{alg:createTime}
    \begin{algorithmic}[1]
    \Procedure{createTime}{$knownDeclarations$, $decl$}
      \If{decl $\in$ knownDeclarations}
        \ForEach{observation \(\in\) knownDeclarations[decl]}
          \If{observation.kind $is$ ADDED}
            \State \Return observation.time
          \EndIf
        \EndFor
      \EndIf
      \State \Return 0
    \EndProcedure
  \end{algorithmic}
\end{algorithm}

\begin{algorithm}
  \scriptsize
  \caption{Determining if a declaration is ignored.}
  \label{alg:ignored}
    \begin{algorithmic}[1]
    \Procedure{ignored}{$knownDeclarations$, $decl$}
      \If{decl $\in$ knownDeclarations}
        \ForEach{observation \(\in\) knownDeclarations[decl]}
          \If{observation.kind $is$ CHANGED}
            \State \Return \textsc{True}
          \EndIf
        \EndFor
      \EndIf
      \State \Return \textsc{False}
    \EndProcedure
  \end{algorithmic}
\end{algorithm}

\noindent We then process the stack of commits as follows:

\begin{itemize}
  \item When a declaration is introduced, we store the location, type inference status, and the commit timestamp (lines 8-10).
  \item When a declaration's type inference status changes (shown in \Cref{alg:initialInferred}), we compute the difference in time between the present commit and the initial commit (shown in \Cref{alg:createTime}), and output location information, type inference status, time to change, and type of outcome (\texttt{CHANGED}) (line 14), and then add the location to an ignore list as shown in \Cref{alg:ignored} (since we only track time to first change).
  \item When a declaration disappears (i.e., it is no longer found in the code, or the file is deleted), we perform the same calculations, outputting with the relevant outcome type (\texttt{REMOVED} (line 17) or \texttt{FILE\_DELETED} (not shown, but treats every declaration as removed) respectively).
  \item Finally, if the commit is the final commit, all remaining declarations in the repository are processed as above, with the outcome of \texttt{OBSERVATION\_ENDED} (lines 22-28).
\end{itemize}

\subsection{Analysis}
\label{sec:analysis-methods}

We consider the frequency of type inference, per project, in \ref*{rq:usage} as both a percentage of inferred declarations for a given location and a percentage of inferred declarations at each location based on the total number of inferrable locations in the project.
This allows us to describe differences in usage between locations and allows us to consider how each location factors into overall usage frequency.

We then examine the relationships between project size metrics and project-level usage of type inference using Pearson's \(r\)~\citep{pearson00}.
We first verify there are no higher-order relationships between any of the project size metrics and type inference usage by examining scatter plots (available in the replication package~\citep{flint21:_replic_packag_inves_type_infer_usage_kotlin}).

For \ref*{rq:context}, we look deeper into the usage of type inference, examining some specific contexts where type inference is used.
First, we collect initializer expressions of inferred variable declarations, examining how frequently different kinds of expressions are used.
Next, we examine how frequently method calls in initializers are to file-local methods and functions.
Then, as testing is a different activity from general software development~\citep{spadini18,bertolino07}, we look for differences in testing code by repeating our \ref*{rq:usage} analysis on only testing-related files.
Finally, as Kotlin and Java are interoperable, with the latter having different rules for type inference, we again repeat our \ref*{rq:usage} analysis but on only mixed-language projects to see if the presence of Java affected inference use in Kotlin code.
We also perform the Kruskal-Wallis \(H\) test to verify differences and report the \(p\) values. This test is used because of the censored, non-normal nature of the data analyzed.

For \ref*{rq:over-time} we need to perform a survival analysis.
Survival analysis is a method by which we can quantify and analyze the time to an event from an initial state, particularly in the presence of missing data about the time to the event (such as not knowing if the event has occurred due to attrition)~\citep{kleinbaum12:_surviv_analy}.
As this sort of censoring is likely in our dataset, this analysis is more useful than simply presenting mean-time-to-change.
Moreover, survival curves like the Kaplan-Meier estimator~\citep{borgan14:_kaplan_meier_estim} include more information than an average, allowing for richer analysis.

\begin{table*}[ht]
  \centering
  \caption{Usage of type inference in Kotlin by location, as percentages of specific locations in a project (outliers included; min \& max omitted as equal).}
  \label{tab:type-inf-place}
  \small
  \renewcommand{\tabcolsep}{0.5em}
\begin{tabular}{llrrrrrrr}
  \toprule
   \textbf{Location} & \textbf{Inferred?} & \textbf{mean} & \textbf{std} & \textbf{min} & \textbf{25\%} & \textbf{50\%} & \textbf{75\%} & \textbf{max} \\
  \midrule
  \multirow[t]{2}{*}{\textbf{Field}} & \textbf{Inferred} & 26.95 & 25.18 & 0 & 7.14 & 22.00 & 38.89 & 100 \\
  \textbf{} & \textbf{Not Inferred} & 73.05 & 25.18 & 0 & 61.11 & 78.00 & 92.86 & 100 \\
  \multirow[t]{2}{*}{\textbf{Global Variable}} & \textbf{Inferred} & 87.20 & 28.61 & 0 & 100.00 & 100.00 & 100.00 & 100 \\
  \textbf{} & \textbf{Not Inferred} & 12.80 & 28.61 & 0 & 0.00 & 0.00 & 0.00 & 100 \\
  \multirow[t]{2}{*}{\textbf{Lambda Arg}} & \textbf{Inferred} & 92.30 & 22.41 & 0 & 100.00 & 100.00 & 100.00 & 100 \\
  \textbf{} & \textbf{Not Inferred} & 7.70 & 22.41 & 0 & 0.00 & 0.00 & 0.00 & 100 \\
  \multirow[t]{2}{*}{\textbf{Local Variable}} & \textbf{Inferred} & 86.46 & 19.22 & 0 & 80.00 & 94.12 & 100.00 & 100 \\
  \textbf{} & \textbf{Not Inferred} & 13.54 & 19.22 & 0 & 0.00 & 5.88 & 20.00 & 100 \\
  \multirow[t]{2}{*}{\textbf{Loop Var}} & \textbf{Inferred} & 97.92 & 12.83 & 0 & 100.00 & 100.00 & 100.00 & 100 \\
  \textbf{} & \textbf{Not Inferred} & 2.08 & 12.83 & 0 & 0.00 & 0.00 & 0.00 & 100 \\
  \multirow[t]{2}{*}{\textbf{Return Type}} & \textbf{Inferred} & 52.05 & 38.76 & 0 & 11.11 & 50.00 & 100.00 & 100 \\
  \textbf{} & \textbf{Not Inferred} & 47.95 & 38.76 & 0 & 0.00 & 50.00 & 88.89 & 100 \\
  \bottomrule
  \end{tabular}
\end{table*}

\begin{figure}[ht]
  \centering
  \includegraphics[width=\linewidth]{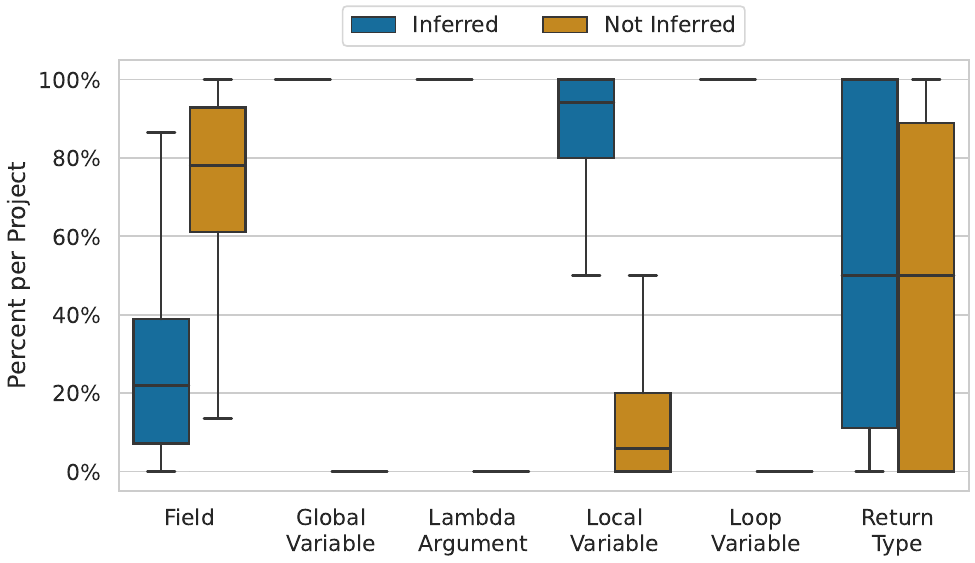}
  \caption{Usage of type inference per inferable location (medians of each category sum to roughly 100\%), outliers not shown.  See also \Cref{tab:type-inf-place}.}
  \label{fig:usage-total-frequency}
\end{figure}

\begin{figure}[ht]
  \centering
  \includegraphics[width=\linewidth]{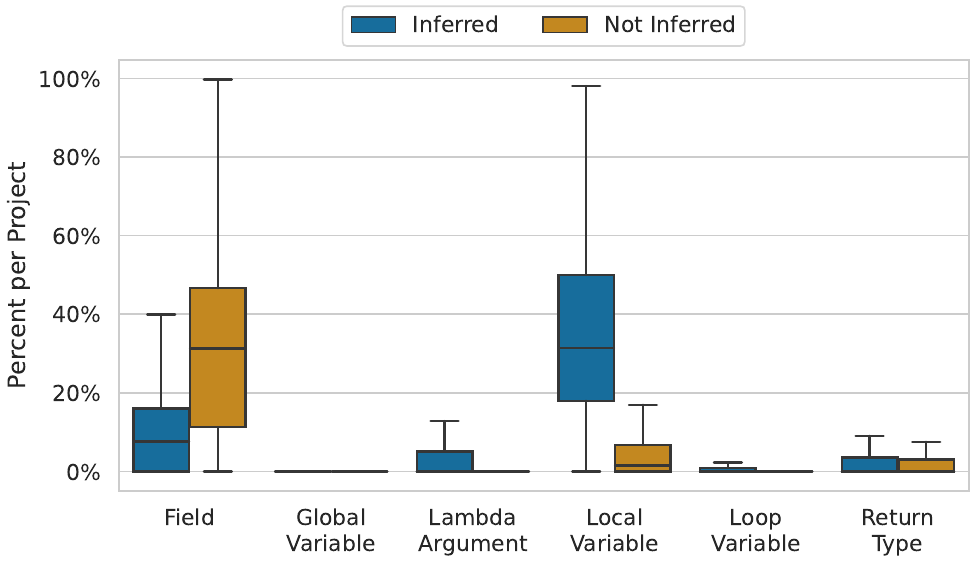}
  \caption{Usage of type inference (as percentage) in all locations of a project (all medians sum to roughly 100\%), outliers not shown.}
  \label{fig:usage-place}
\end{figure}

For data visualization and statistical summarization, we use the \texttt{pandas} library~\citep{team20}.
We measure the strength and direction of relationships between type inference usage and various project size metrics using Pearson's \(r\)~\citep{pearson00}, as implemented by the \texttt{scipy} library~\citep{virtanen20:_scipy}.
For our survival analysis, we use the Kaplan-Meier estimator~\citep{borgan14:_kaplan_meier_estim}, calculated using the \texttt{lifelines} library~\citep{davidson-pilon19}.

\section{Results}
\label{sec:results}

In this section, we answer the research questions via the results.

\subsection{\ref*{rq:usage}: How often do Kotlin developers use type inference?}
\label{sec:rq-usage-results}

The first question investigates where and how often developers use type inference in Kotlin, where allowed.
We show our analysis via two plots.
The first of these plots, \Cref{fig:usage-total-frequency}, shows how often, for a given location, type inference is used (that is, the medians for each pair of bars will sum roughly to 100\%, e.g. for Fields the medians are around 22\% and 78\%).  This is useful to see if a given location (fields, globals, locals, etc.) is more likely to be inferred or not.
The second plot, \Cref{fig:usage-place}, looks at all locations in a given project, and shows how often type inference is used in each location, relative to the total number of locations in a given project.  Here, the sums of the medians across all locations is roughly 100\%.  This is useful to see, for a single project, what locations appear more often and if those are inferred or not.
Both plots have had outliers removed, and a full statistical summary (including outliers) for \Cref{fig:usage-total-frequency} is shown in \Cref{tab:type-inf-place}.
We discuss these results in more detail.

First, we consider per-location percentages, shown in \Cref{fig:usage-total-frequency}.
In this figure, we see that some locations use type inference significantly more than others: in particular, 94\% of local variables use type inference in the median project.
Notably, lambda arguments are inferred 100\% of the time, even in the 25\textsuperscript{th} percentile: the usage of type annotations in lambda arguments is quite rare.
This pattern is seen for loop variables as well.
However, it appears that fields are inferred in only about one-fourth of instances.

\begin{finding}{}{fd:kotlin-relative-usage}
  In Kotlin, lambda arguments, global variables, loop variables, and local variables almost always have their types inferred.
  Fields are often not inferred, similar to a prior study~\citep{souza14:_how}.
  Function return types are evenly split.
\end{finding}

Next, when we look at the median values in \Cref{fig:usage-place}, we see that about 39\% of type-inference-capable declarations (median lines in the blue boxes, 7.58\%+0\%+0\%+31.43\%+0\%+0\%=39.01\%) use type inference.
This means type inference is a popular feature, and the differences that we see in \Cref{fig:usage-total-frequency} are reflected in the project-level data.

\begin{finding}{}{fd:usage-generally}
  Type inference is a popular feature in Kotlin, with a median of 42\% of declarations in a project using it.
\end{finding}

\begin{table}[htbp]
  \centering
  \caption{Correlation (Pearson's \(r\)) between overall usage of type inference and project size variables.}
  \label{tab:correlation-size}
  \small
\begin{tabular}{lrr}
  \toprule
   & \textbf{r} & \textbf{p} \\
  \midrule
  \textbf{Number of Files} & -0.07 & 0.00 \\
  \textbf{Number of Statements} & -0.02 & 0.00 \\
  \textbf{Number of Stars} & 0.00 & 0.14 \\
  \textbf{Number of Developers} & 0.00 & 0.44 \\
  \bottomrule
  \end{tabular}
\end{table}

Additionally, we consider the correlation between several project size metrics and overall usage of type inference (see \Cref{tab:correlation-size}).
Overall, we see that while several correlations can be said to exist (having \(p\)-values less than 0.05), the correlations detected are small and negative.
This is to say, it appears unlikely that various project size metrics are related to the use of type inference.

\begin{finding}{}{fd:no-correlation}
  Correlations (where they exist) between percentage of type-inferred declarations and project size metrics are incredibly small, that is, there are no strong relations between project size and the use of type inference.
\end{finding}

\subsection{\ref*{rq:context}: Is type inference used differently in different contexts?}
\label{sec:context-results}

Since the last research question showed the use of type inference varies based on location, in this section we look a bit deeper at how type inference is used in different contexts such as initializer expressions, (im)mutable declarations, testing code, and mixed-language projects.

\subsubsection{\ref*{rq:rhs}: Is type inference used more frequently with certain kinds of initializer expressions?}
\label{sec:rhs-results}

\begin{figure}[ht]
  \centering
    \includegraphics[width=0.9\linewidth]{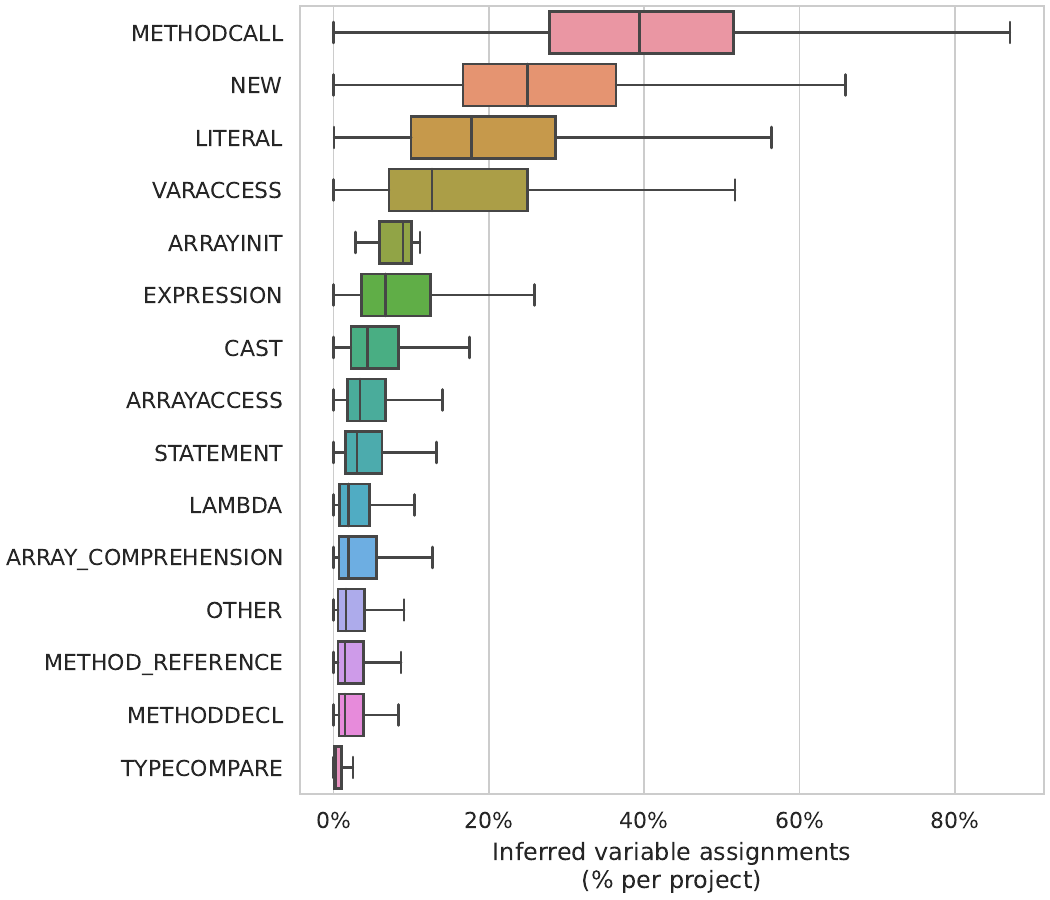}
  \caption{Most common expression types in variable initializers (per project).}
  \label{fig:rhs-results}
\end{figure}

To better understand the behavior of developers with respect to type inference, we consider the kind of expression on the right-hand side (RHS) of variable declarations (\ie, the initializer).
The results are shown in \Cref{fig:rhs-results}.
This includes all variable declarations that allow initializers, such as locals, globals, and fields.

The most common kind of RHS expression is a method call, with almost 40\% of all RHS expressions, followed by object instantiations (including \lstinline[language=kotlin,style=default]|var foo = object {};|), literals, and variable accesses.
In fact, these four kinds account for over 85\% of all RHS expressions in the median project.

\begin{finding}{}{fd:most-common-method-call}
  Method calls, followed by object instantiations, are the most common initializers for type-inferred variables.
\end{finding}

Given that method calls are the most common initializer for inferred variables, understanding if calls are to methods local to the module or not is important since local calls can be analyzed modularly.
We counted the number of non-file-local calls and possibly file-local calls and show these results in \Cref{tab:rhs-calls}.
The results show that calls are overwhelmingly to non-file-local methods.

\begin{finding}{}{fd:rhs-non-local}
  Most method calls in inferred variable initializers are to methods that are not local to the file.
\end{finding}

To summarize, for initializer expressions of type-inferred declarations, the most common kind of expression is a non-file-local method call.

\begin{table}[ht]
    \centering
    \caption{Are method calls used for inference to local methods? (as defined in \Cref{sec:measurements})}
    \label{tab:rhs-calls}
  \small
\begin{tabular}{lr}
  \toprule
   & \textbf{Number of Calls} \\
  \midrule
  \textbf{Not Local} & 8,735,180 \\
  \textbf{Possibly File-Local} & 871,715 \\
  \bottomrule
  \end{tabular}
\end{table}

\subsubsection{\ref*{rq:mutability}: Is the use of type inference different for immutable variables or mutable variables?}
\label{sec:rq-mutability-results}

\begin{figure*}[htbp]
  \centering
  \begin{subfigure}[t]{\linewidth}
    \centering
    \includegraphics[width=0.8\linewidth]{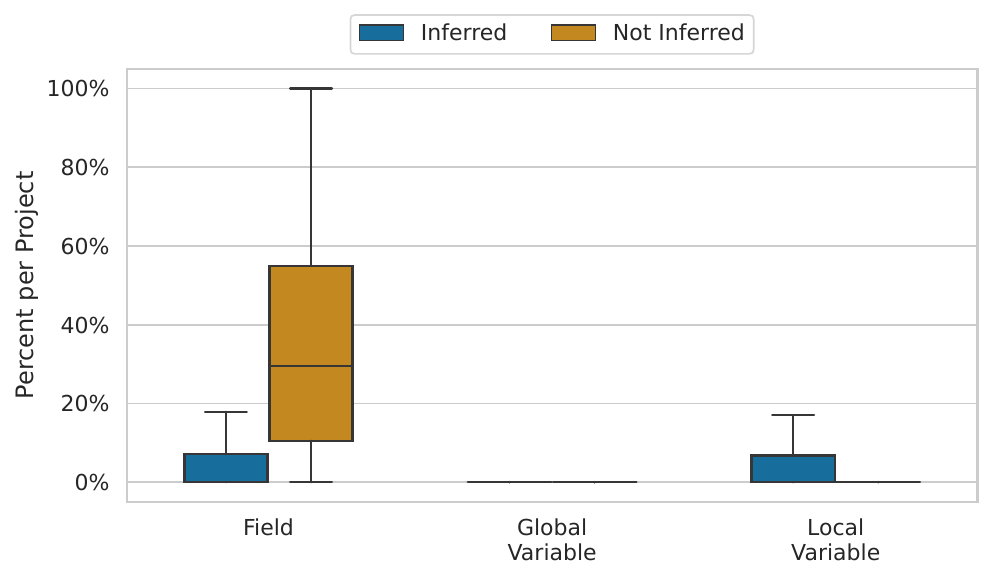}
    \caption{Per-project percent of type inference usage by location when considering only immutable variables.}
    \label{fig:mutability-immutable}
  \end{subfigure}
  % \hspace*{3em}
  \begin{subfigure}[t]{\linewidth}
    \centering
    \includegraphics[width=0.8\linewidth]{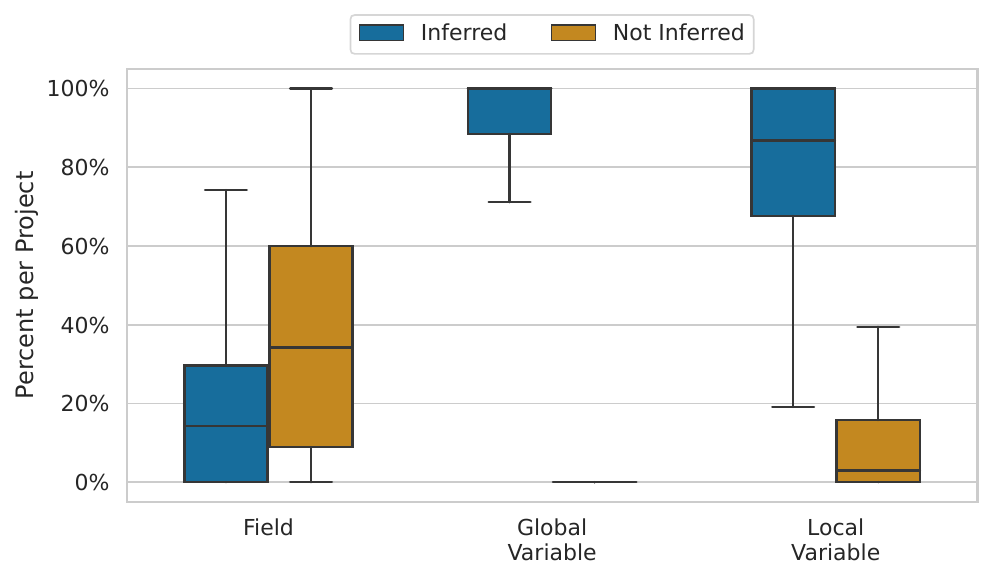}
    \caption{Per-project percent of type inference usage by location when considering only mutable variables.}
    \label{fig:mutability-mutable}
  \end{subfigure}
  \caption{Per-project percent of type inference usage by location and mutability (outliers not shown).}
  \label{fig:mutability}
\end{figure*}

Next, we consider if variable mutability affects the usage of type inference.
The results are shown in \Cref{fig:mutability,tab:mutability}.

\begin{table*}[htbp]
  \centering
  \caption{How often variables are inferred, per project, when taking into account place and mutability status (outliers included).}
  \label{tab:mutability}
  \small
  \renewcommand{\tabcolsep}{0.15em}
\begin{tabular}{lllrrrrrrr}
  \toprule
   \textbf{Location} & \textbf{Inferred?} & \textbf{Mutable?} & \textbf{mean} & \textbf{std} & \textbf{min} & \textbf{25\%} & \textbf{50\%} & \textbf{75\%} & \textbf{max} \\
  \midrule
  \multirow[t]{4}{*}{\textbf{Field}} & \multirow[t]{2}{*}{\textbf{Inferred}} & \textbf{Mutable} & 20.12 & 22.17 & 0 & 0.00 & 14.29 & 29.69 & 100 \\
  \textbf{} & \textbf{} & \textbf{Not Mutable} & 6.82 & 14.73 & 0 & 0.00 & 0.00 & 7.14 & 100 \\
  \textbf{} & \multirow[t]{2}{*}{\textbf{Not Inferred}} & \textbf{Mutable} & 37.34 & 30.20 & 0 & 8.89 & 34.21 & 60.00 & 100 \\
  \textbf{} & \textbf{} & \textbf{Not Mutable} & 35.72 & 30.09 & 0 & 10.53 & 29.58 & 54.98 & 100 \\
  \midrule
  \multirow[t]{4}{*}{\textbf{Global Variable}} & \multirow[t]{2}{*}{\textbf{Inferred}} & \textbf{Mutable} & 83.99 & 31.82 & 0 & 88.46 & 100.00 & 100.00 & 100 \\
  \textbf{} & \textbf{} & \textbf{Not Mutable} & 3.21 & 14.73 & 0 & 0.00 & 0.00 & 0.00 & 100 \\
  \textbf{} & \multirow[t]{2}{*}{\textbf{Not Inferred}} & \textbf{Mutable} & 8.49 & 23.70 & 0 & 0.00 & 0.00 & 0.00 & 100 \\
  \textbf{} & \textbf{} & \textbf{Not Mutable} & 4.32 & 17.26 & 0 & 0.00 & 0.00 & 0.00 & 100 \\
  \midrule
  \multirow[t]{4}{*}{\textbf{Local Variable}} & \multirow[t]{2}{*}{\textbf{Inferred}} & \textbf{Mutable} & 79.34 & 24.24 & 0 & 67.65 & 86.79 & 100.00 & 100 \\
  \textbf{} & \textbf{} & \textbf{Not Mutable} & 7.12 & 15.70 & 0 & 0.00 & 0.00 & 6.82 & 100 \\
  \textbf{} & \multirow[t]{2}{*}{\textbf{Not Inferred}} & \textbf{Mutable} & 10.93 & 17.13 & 0 & 0.00 & 2.95 & 15.79 & 100 \\
  \textbf{} & \textbf{} & \textbf{Not Mutable} & 2.61 & 8.39 & 0 & 0.00 & 0.00 & 0.00 & 100 \\
  \bottomrule
  \end{tabular}
\end{table*}

The most notable difference is in global variables, which, as we observed previously, are almost always inferred.
Here, we see that global variables are also almost always mutable, with 100\% of global variables being mutable and inferred in the median project (statistically significant, with \(p < 0.001\)).

Local variables, however, exhibit more variation with respect to mutability.
While local variables are by and large inferred, a small amount (about 10\%) are immutable, with most of the immutable local variables being inferred (statistically significant, with \(p < 0.001\)).

Finally, most fields are mutable (about 57\%), no matter the inference status.
However immutable non-inferred fields are more common than mutable inferred fields (statistically significant, with \(p < 0.001\)).

\begin{finding}{}{fd:mutability-summary}
  Overall, we find that across the three categories of declarations, more declarations are mutable.
  Additionally, when variables are immutable, inference is used in a similar pattern to usage in mutable declarations: more immutable fields are not inferred than are inferred and more local variables are inferred than not.
\end{finding}

\subsubsection{\ref*{rq:tests}: Is the usage of type inference different in testing classes?}
\label{sec:results-tests}

\begin{figure*}[htbp]
  \centering
  \begin{subfigure}[t]{\linewidth}
    \centering
    \includegraphics[width=0.8\linewidth]{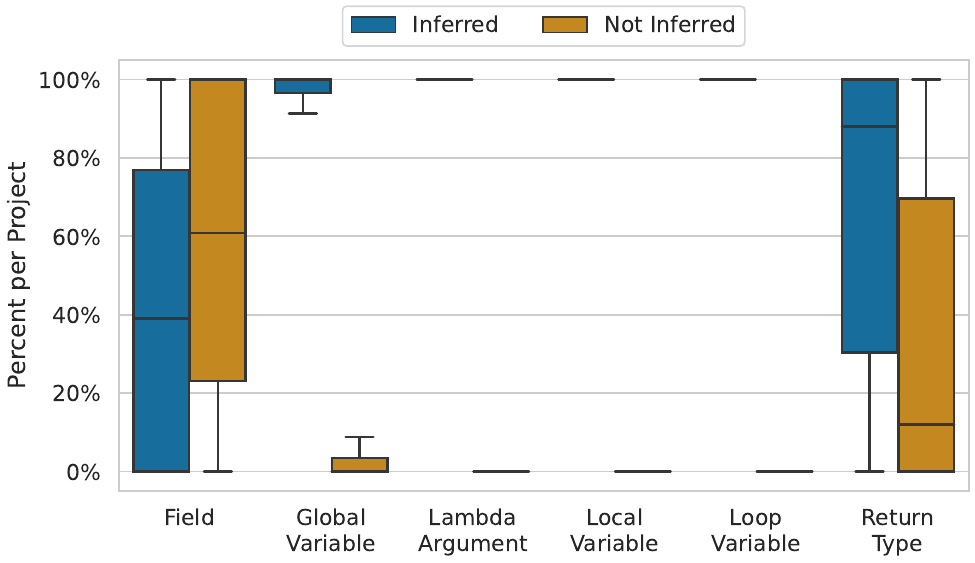}
    \caption{Usage of type inference by location in testing code.}
    \label{fig:usage-in-tests}
  \end{subfigure}
  \begin{subfigure}[t]{\linewidth}
    \centering
    \includegraphics[width=0.8\linewidth]{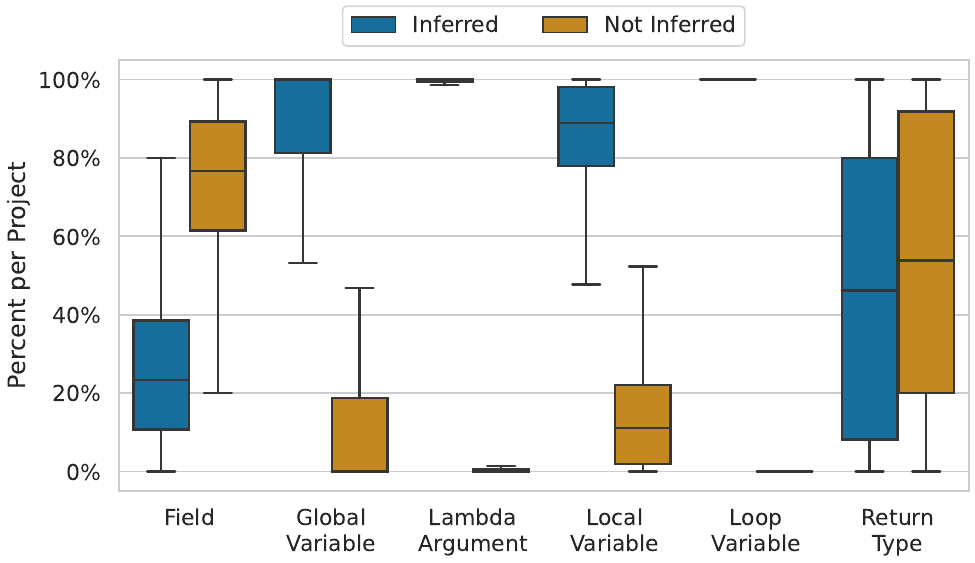}
    \caption{Usage of type inference by location in mixed projects.}
    \label{fig:usage-mixed}
  \end{subfigure}
  \caption{Per project percent of type inference usage by location, for test classes (\ref*{fig:usage-in-tests}) and mixed projects (\ref*{fig:usage-mixed}).
    (Compare to \Cref{fig:usage-total-frequency}.)}
  \label{fig:contexts}
\end{figure*}

Next, we look at the use of type inference in testing code, shown in \Cref{fig:usage-in-tests,tab:usage-in-tests}.
When we compare these results to \Cref{tab:type-inf-place}, we see the use of type inference in testing code is different than previously observed.
Specifically, we see two locations with substantially different usage patterns: fields and return types.

Fields see type inference used 17\% more often in testing classes (statistically significant, with \(p < 0.001\)).
Note, however, even in testing classes, fields are still annotated about 60\% of the time.
Return types, on the other hand, see quite a bit higher usage in testing code, with median projects having 88\% of their instances inferred as opposed to 50\% shown in \Cref{tab:type-inf-place}, (statistically significant, with \(p < 0.001\)).

\begin{finding}{}{fd:test-usage}
  Testing code seems to use type inference more frequently, for field declarations and function return types.
\end{finding}

\begin{table*}[htbp]
  \centering
  \caption{Usage of inference in testing code per project by allowed location (outliers shown).
    Median values marked \hlover{green} are higher than similar values in \Cref{tab:type-inf-place}.
    Median values marked \hlunder{orange} are lower than similar values in \Cref{tab:type-inf-place}.}
  \label{tab:usage-in-tests}
  \small
  \renewcommand{\tabcolsep}{0.45em}
\begin{tabular}{llrrrrrrr}
  \toprule
   \textbf{Location} & \textbf{Inferred?} & \textbf{mean} & \textbf{std} & \textbf{min} & \textbf{25\%} & \textbf{50\%} & \textbf{75\%} & \textbf{max} \\
  \midrule
  \multirow[t]{2}{*}{\textbf{Field}} & \textbf{Inferred} & 43.77 & 37.55 & 0 & 0.00 & \hlover{39.10} & 76.92 & 100 \\
  \textbf{} & \textbf{Not Inferred} & 56.23 & 37.55 & 0 & 23.08 & \hlunder{60.90} & 100.00 & 100 \\
  \multirow[t]{2}{*}{\textbf{Global Variable}} & \textbf{Inferred} & 86.22 & 29.99 & 0 & 96.49 & 100.00 & 100.00 & 100 \\
  \textbf{} & \textbf{Not Inferred} & 13.78 & 29.99 & 0 & 0.00 & 0.00 & 3.51 & 100 \\
  \multirow[t]{2}{*}{\textbf{Lambda Arg}} & \textbf{Inferred} & 91.25 & 24.85 & 0 & 100.00 & 100.00 & 100.00 & 100 \\
  \textbf{} & \textbf{Not Inferred} & 8.75 & 24.85 & 0 & 0.00 & 0.00 & 0.00 & 100 \\
  \multirow[t]{2}{*}{\textbf{Local Variable}} & \textbf{Inferred} & 98.52 & 8.25 & 0 & 100.00 & 100.00 & 100.00 & 100 \\
  \textbf{} & \textbf{Not Inferred} & 1.48 & 8.25 & 0 & 0.00 & 0.00 & 0.00 & 100 \\
  \multirow[t]{2}{*}{\textbf{Loop Var}} & \textbf{Inferred} & 98.96 & 9.45 & 0 & 100.00 & 100.00 & 100.00 & 100 \\
  \textbf{} & \textbf{Not Inferred} & 1.04 & 9.45 & 0 & 0.00 & 0.00 & 0.00 & 100 \\
  \multirow[t]{2}{*}{\textbf{Return Type}} & \textbf{Inferred} & 66.24 & 40.10 & 0 & 30.30 & \hlover{88.00} & 100.00 & 100 \\
  \textbf{} & \textbf{Not Inferred} & 33.76 & 40.10 & 0 & 0.00 & \hlunder{12.00} & 69.70 & 100 \\
  \bottomrule
  \end{tabular}
\end{table*}

\subsubsection{\ref*{rq:mixed-projects}: Is the usage of type inference different in Kotlin projects containing both Kotlin and Java source files?}
\label{sec:results-mixed-projects}

Finally, we look at projects that also contain Java source files.
This data is a subset of the overall data, and only used in this sub-section.
Of the projects studied, 12.32\% include Java source files.
The median project including Java source files had \(7.69\%\) of the code in Java (\(\text{mean} = 12.96\%, \text{std} = 14.074\)).
In \Cref{fig:usage-mixed,tab:usage-mixed} we see type inference usage in mixed Kotlin-Java projects.
These minor differences could be explained by the subsetting effect of partitioning at the project level.
However, these differences are all statistically significant, with \(p < 0.001\).

When comparing these results to \Cref{tab:type-inf-place}, we see three locations that differ in their usage of type inference: fields, local variables, and return types.
Fields were actually inferred more in mixed projects, while local variables and return types were inferred less in mixed projects.
However the differences were all very small, ranging from 1.29\%-5.23\%.

\begin{finding}{}{fd:usage-mixed}
    Even in the presence of Java code, type inference seems to be used similarly to its use in all projects, where the differences are overall quite small ($\sim$5\%).
\end{finding}

\begin{table*}[htbp]
  \centering
  \caption{Usage of inference in mixed Kotlin-Java projects by allowed location (outliers shown).
    Median values marked \hlover{green} are higher than similar values in \Cref{tab:type-inf-place}.
    Median values marked \hlunder{orange} are lower than similar values in \Cref{tab:type-inf-place}.}
  \label{tab:usage-mixed}
  \small
  \renewcommand{\tabcolsep}{0.45em}
\begin{tabular}{llrrrrrrr}
  \toprule
   \textbf{Location} & \textbf{Inferred?} & \textbf{mean} & \textbf{std} & \textbf{min} & \textbf{25\%} & \textbf{50\%} & \textbf{75\%} & \textbf{max} \\
  \midrule
  \multirow[t]{2}{*}{\textbf{Field}} & \textbf{Inferred} & 26.94 & 21.80 & 0 & 10.67 & \hlover{23.29} & 38.46 & 100 \\
  \textbf{} & \textbf{Not Inferred} & 73.06 & 21.80 & 0 & 61.54 & \hlunder{76.71} & 89.33 & 100 \\
  \multirow[t]{2}{*}{\textbf{Global Variable}} & \textbf{Inferred} & 82.63 & 31.71 & 0 & 81.25 & 100.00 & 100.00 & 100 \\
  \textbf{} & \textbf{Not Inferred} & 17.37 & 31.71 & 0 & 0.00 & 0.00 & 18.75 & 100 \\
  \multirow[t]{2}{*}{\textbf{Lambda Arg}} & \textbf{Inferred} & 91.60 & 21.29 & 0 & 99.42 & 100.00 & 100.00 & 100 \\
  \textbf{} & \textbf{Not Inferred} & 8.40 & 21.29 & 0 & 0.00 & 0.00 & 0.58 & 100 \\
  \multirow[t]{2}{*}{\textbf{Local Variable}} & \textbf{Inferred} & 84.05 & 18.90 & 0 & 77.92 & \hlunder{88.89} & 98.09 & 100 \\
  \textbf{} & \textbf{Not Inferred} & 15.95 & 18.90 & 0 & 1.91 & \hlover{11.11} & 22.08 & 100 \\
  \multirow[t]{2}{*}{\textbf{Loop Var}} & \textbf{Inferred} & 98.20 & 11.21 & 0 & 100.00 & 100.00 & 100.00 & 100 \\
  \textbf{} & \textbf{Not Inferred} & 1.80 & 11.21 & 0 & 0.00 & 0.00 & 0.00 & 100 \\
  \multirow[t]{2}{*}{\textbf{Return Type}} & \textbf{Inferred} & 46.62 & 36.69 & 0 & 8.11 & \hlunder{46.15} & 80.00 & 100 \\
  \textbf{} & \textbf{Not Inferred} & 53.38 & 36.69 & 0 & 20.00 & \hlover{53.85} & 91.89 & 100 \\
  \bottomrule
  \end{tabular}
\end{table*}

\subsection{\ref*{rq:over-time}: Does the use of type inference change over time?}
\label{sec:over-time-results}

\begin{figure}[htbp]
  \centering
    \includegraphics[width=\linewidth]{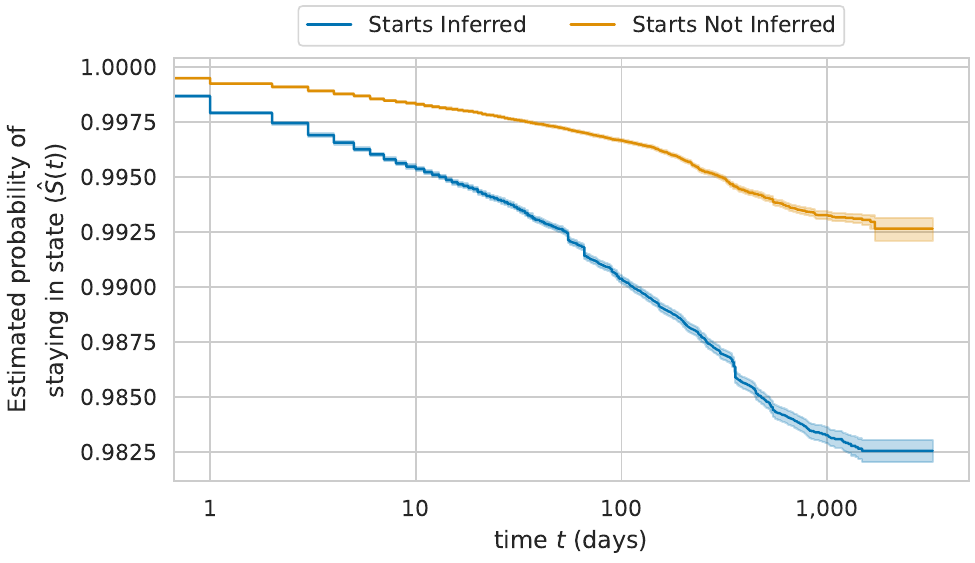}
  \caption{How long does a variable keep its annotation status?  Note that the time axis is shown as logarithmic for space.}
  \label{fig:survival-status}
\end{figure}

For the last research question, we consider how type inference behavior changes over time, using a survival analysis with the Kaplan-Meier estimator.
We show the results in \Cref{fig:survival-status}, where the lines are the probability of inference status surviving that long and the shaded areas represent confidence bounds.
For example, the blue line represents the probability that a location using type inference would keep using it over the next 1,000 days.

Overall we see the probability of survival is \emph{high}, with the probability at \(> 1,000\) days being greater than 98\% in both cases.
Of note, the probability that a type annotation is removed is incredibly low, with a 0.75\% chance after 1,000 days.
Similarly, the probability that a type annotation is added is low, at 1.75\%.

\begin{finding}{}{fd:vars-stay-in-state}
  Variable declarations do not tend to add or remove type annotations at a later date.
\end{finding}

\section{Threats to Validity}
\label{sec:threats-validity}

In this section, we discuss threats to the construct, internal, external, and conclusion validity of our study.

\subsection{Construct Validity}
\label{sec:construct-threats}

Our heuristic for determining object instantiations versus method calls in Kotlin (used in \ref*{rq:rhs}) may be somewhat lossy, particularly for star imports.
Our fallback strategy of looking at the case of the first letter may not be perfect, but standard naming conventions (pascal-case class names and camel-case method names) mean it should be accurate if developers follow those standard conventions.
A brief manual analysis of our heuristic showed a 97.6\% correctness (on a sample of 385 instances, for a confidence level of 95\% with 5\% margin of error on an initial 14,4395,555 instance population), and a Cohen's Kappa of \(\kappa = .9318\) between the human rater and the heuristic. Thus, this threat is minimal.

Correct detection of file-local calls requires complete type information, which is not available in the Boa system.
Instead, our heuristic uses incomplete information and may overestimate the number of file-local calls.
Since there are ten times as many non-local calls as possibly-file-local calls, this particular threat is minimized.

Detection of testing files may miss tests not including the word \enquote{test} in their path and not using one of the listed testing libraries, though this technique has been utilized in prior works~\citep{nakamaru20,keshk23}.

\subsection{Internal Validity}
\label{sec:internal-threats}

As an exploratory study that focuses on characterizing the usage of type inference by developers in Kotlin, without attributing causality, we have few threats to internal validity.
However, the reasons why developers choose to use type inference in different places are unclear.
As we do not have access to the developers, we are unable to discover their reasoning or accurately determine if types were automatically annotated by tooling without developer intervention.
That is, our characterizations are of artifacts, not behavior itself.

\subsection{External Validity}
\label{sec:external-threats}

In this study, there are three major barriers to external validity.
First is that this study considers only Kotlin, and generalization to other statically typed languages allowing type inference may not be possible.
Other languages have differing rules for type inference, and may not have allowed it at the locations or similar time frames.

Additionally, as we only study open-source GitHub repositories accessible during August 2021, our results may not apply to proprietary code or code written in newer versions of Kotlin.

Finally, we do not filter out toy projects: while this means we describe general behavior, our results may present some skew because of it.
However, as we note in \Cref{fd:no-correlation}, as we observed no relationship between type inference usage and various project size metrics, this threat is minimized.

\subsection{Conclusion Validity}
\label{sec:conclusion-validity}

Because the research questions are about general patterns in the use of type inference, presenting measures of central tendency is appropriate.
The use of Pearson's \(r\) to determine if there are relationships between measures of project size and usage of type inference is supported by the nature of the data.
Similarly, the use of the Kaplan-Meier estimator is appropriate as we have right-censored data with respect to the amount of time a variable stays in its initial inference state.

\section{Discussion}
\label{sec:discussion}

Given our findings, we now consider their implications for practitioners, language designers, tool developers, and researchers.

\subsection{Usage of Type Inference (Findings 1/3/6/7/8)}
\label{sec:usage-type-inference}

In these findings, we see that type inference is a popular feature in Kotlin, and its usage does not change with project characteristics like size, popularity, or number of developers.
However, we see that different development contexts may influence where type inference is used and the extent of use.
Testing code, for instance, uses type inference more frequently in field declarations and function return types than in other places.

For researchers seeking to understand the cognitive impact of using type annotations, the patterns observed in this study may help to provide initial evidence for future studies.
Since declarations are annotated more often in some places, this suggests that developers have different information needs and further research is necessary.
This is especially clear when we consider the difference between testing and non-testing code: there appears to be a categorical difference which may cause developers to not annotate fields and function return types.

Finally, as type inference appears to be a frequently used feature, practicing developers may want to more carefully consider when and why they use this feature, particularly given work such as \citet{spiza14:_type_apis}, that found type annotations alone (\ie, without type checking) improve the usability of APIs.

\subsection{Initializers Call Non-Local Methods (Findings 4/5)}
\label{sec:initializer-call-non}

With non-local method calls as the most common variable initializer, we see a possibility for the design of clever type checkers and type inference engines that save information in more readily accessible or usable forms to speed up their execution.
If object instantiation or literal expressions were more common, then this may not be necessary, but the need to resolve types may require caching of type information separate from compilation targets.

For tool developers, this knowledge may be helpful in providing information about code that is being read or edited.
While def-use and similar information are already available, improving how and when this information is presented is important.
Moreover, hooking into a language's type inference system may be more important than previously thought.

\subsection{Variables Which Are Inferred Tend To Stay Inferred, and Vice Versa (Finding 9)}
\label{sec:type-inference-usage}

As shown in \Cref{sec:over-time-results}, variables tend to keep their inference status over time.
This result is likely most useful to implementers of languages, as it may be helpful to keep this in mind when implementing compilers: caching results of parts of type-checking information between runs may help to improve type-checking performance, particularly in simpler cases.
The presence of these simpler cases, however, would need to be quantified to better make this case.

This result also shows the endurance of type inference systems in Kotlin.
This may confirm its usability and desirability: even as code changes, the need to annotate (or not) does not appear to change.
Put differently, developers see type inference as helpful while code changes, perhaps because they do not have to re-annotate variables when types change.

Additionally, this result could have implications for tool builders.
If many local variables are not annotated (as suggested by our results), caching type information may be necessary to simplify the implementation of various refactorings or type changes.

Finally, further research is necessary to understand the motivation for adding or removing type annotations in later commits.
In particular, it may be interesting to consider internal factors (\eg, increasing complexity) as well as external factors (such as onboarding new developers) as possible correlates.
\
\section{Related Work}
\label{sec:prior-work}

In this section, we consider related studies on type systems and their usage and language feature mining studies.

\subsection{Type Systems and Usage}
\label{sec:type-system-usage}

\citet{souza14:_how} performed a mining study considering the use of optional typing in Groovy.
They found type annotations were used frequently in formal parameters of constructors and methods (86\% and 100\% medians, respectively), as well as method return types (75\%), however, they were used in less than half of fields (39\%) and few local variables (18\%).
This differs from our findings in Kotlin: fields are more frequently annotated (78\%), but local variables are annotated less frequently (5.9\%).
Next, they consider programmer language experience and its possible effect on usage of type annotations, finding that those with significant statically typed backgrounds are more likely to annotate code,
followed by those with mixed-paradigm backgrounds and those from dynamically typed languages annotating the least.
Finally, they correlate various project size features with type annotation usage, finding that there is no significant correlation with any of these features, including with file-churn.  This is similar to our results.
The main difference is their study was looking at optional typing in dynamic languages while we look at type inference in static languages.
Additionally, we investigate common initializer expression kinds and mutability, as well as looking at the survivability of the type annotations.

\citet{di22:_evolut_type_annot_python} describe type annotation behavior of developers over time in the Python language.
They note that while rare in Python, type annotations are becoming more popular; moreover, type annotations, once in place, are not updated, and they appear not to be removed.
Instead, we consider the more general concept of survival analysis, studying how long type annotation status remains.
We similarly find that type annotations are rarely removed, but also, at least in Kotlin, they are also rarely added.

Other researchers examined the differences between static and dynamic typing in code comprehension for maintenance.
For instance, \citet{mayer12:_empir_study_influen_static_type} found some benefit to developers in using static typing, as did \citet{fischer_hanenberg2015:_empir_inves_effec_type_system}.
This is noticeable in particular when repairing type errors~\citep{kleinschmager12:_do,hanenberg14:_empir_study_impac_static_typin_softw_maint}.
Additionally, \citet{endrikat14:_how_do_api_docum_static} found that the benefits of static typing are strengthened with explicit documentation.
However, \citet{hanenberg10:_exper_static_dynam_type_system} found no statistically significant effect overall on time-to-completion, but found that several subjects using the dynamically-typed language completed tasks faster than others.  This research focused on the benefits or drawbacks of using a specific type system, while our work focuses on how type inference is used in a specific language.
On the other hand, the use of static type systems has been shown to improve the detectability of bugs, at least in JavaScript~\cite{gao17:_to_type_not_type}.
Moreover, \citet{tobinhochstadt_et_al:LIPIcs.SNAPL.2017.17} describe a ten-year retrospective of the addition of type annotations and static typing to the Racket language, describing the evolution of an optional typing system, and some of the effort necessary to start using it.  Our work provides large-scale evidence of the changes in type annotation usage in the Kotlin language, but does not consider the reasons developers may make these changes.

\subsection{Language Feature Mining Studies}
\label{sec:lang-feat-stud}

Prior studies have considered various language features, including in Java.
\citet{dyer14:_misin_ast_java}, for instance, consider how features in JLS2, JLS3, and JLS4 are adopted over time, both before and after release of those versions of the Java language.
One of the features used is a limited form of type inference used in Java-language generic types.
We extend to more instances of type inference, and in one of our research questions, we also consider how the use of type inference evolves over time in Kotlin projects, including consideration of the effects of new language versions on this evolution.

\citet{gois19:_android_kotlin} and \citet{martinez20:_how_android_java_kotlin} studied code quality in Android applications written in Kotlin and the adoption of Kotlin and migration from Java for Android applications, respectively.
To the best of our knowledge, these are the earliest mining studies on Kotlin developer behavior, but they do not focus on the use of type inference in the language.
That said, \citet{mateus_etal20:_adopt_usage_evolut_kotlin_featur_android_devel} found that Android developers using Kotlin do use type inference, and its use has picked up. Our study more thoroughly characterizes the circumstances in which type inference is used by Kotlin developers, and does so in non-Android as well as Android contexts.

\section{Conclusion}
\label{sec:conclusion}

Type inference is a popular language feature for statically typed languages, and in the Kotlin language, a clearly well-used feature.
This suggests that the feature is successful with respect to developer adoption, and may be desirable in other future languages.
Additionally, the initializer for inferred variables is often a method call, and often to a non-local method, potentially increasing the complexity of type inference algorithms.
Kotlin developers seem to use type inference more in testing code, which may require more research to see if that affects code review or maintenance of those tests~\citep{spadini18}.
Finally, variables tend to stay (not) inferred over time: developers rarely see a reason to change variable annotation status.

\section*{Data Availability}

The Boa queries, their output, and all processing scripts are made available in a replication package on Zenodo~\citep{flint21:_replic_packag_inves_type_infer_usage_kotlin}.

\section*{Conflict of Interest}

The authors declare that they have no conflict of interest.

\begin{acknowledgements}
  This work was completed utilizing the Holland Computing Center of the University of Nebraska, which receives support from the Nebraska Research Initiative.
  This work was supported by the U.S. National Science Foundation under the grants CCF-1755890, CCF-2139845, CCF-2124116, and CNS-2346327.
\end{acknowledgements}

\bibliographystyle{spbasic}
\bibliography{bibliography.bib}

\begin{thebibliography}{54}
\bibitem[{Pearson(1900)}]{pearson00}
Pearson K (1900) On the criterion that a given system of deviations from the
  probable in the case of a correlated system of variables is such that it can
  be reasonably supposed to have arisen from random sampling. Philosophical
  Magazine 50:157--175
\end{thebibliography}

\clearpage

\section*{Biographies}

\subsection*{Samuel W. Flint}

\includegraphics[width=1in,height=1.25in,clip,keepaspectratio]{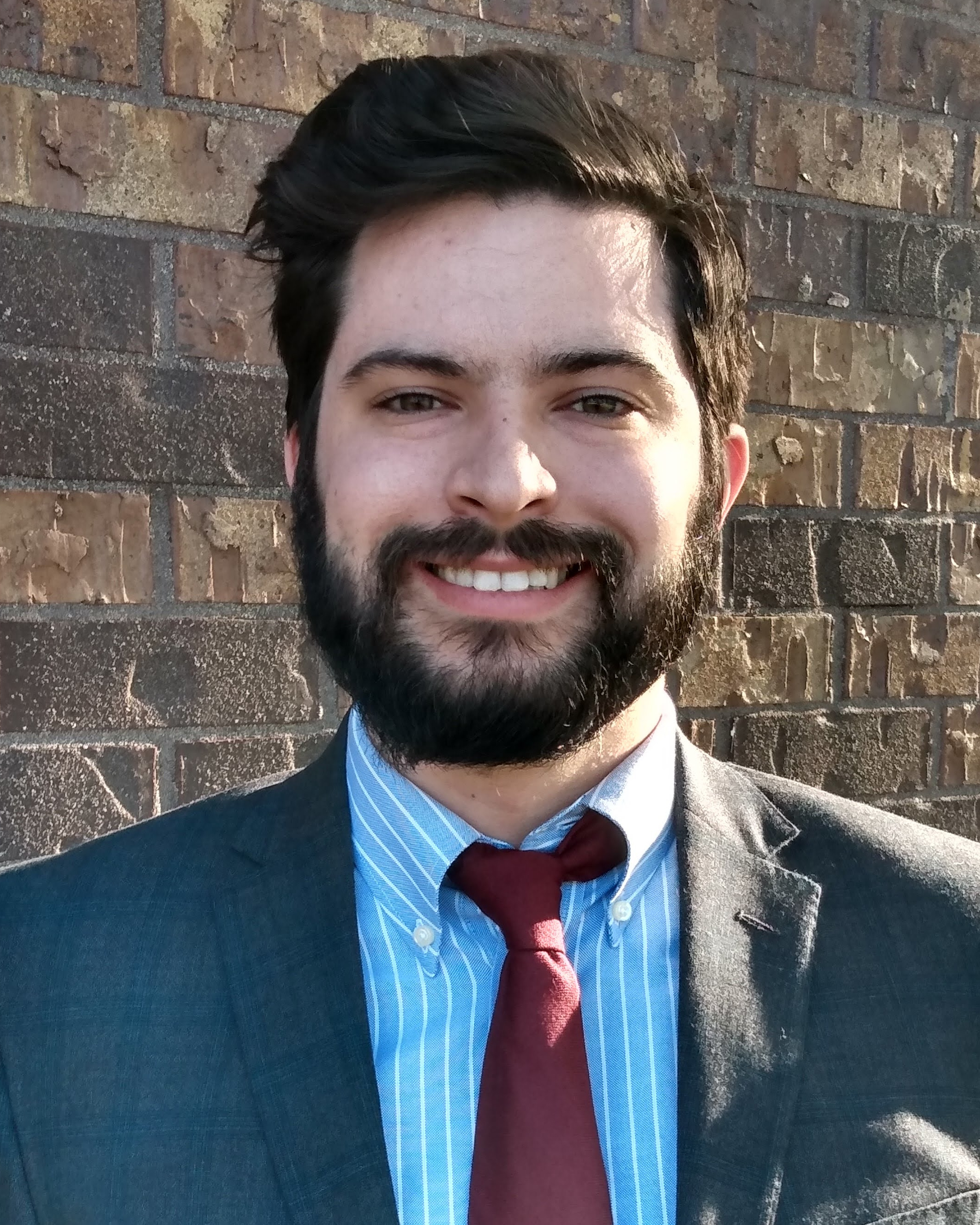}

Samuel W. Flint is a Ph.D. student advised by Robert Dyer at the University of Nebraska--Lincoln, where he also earned his BS degree.  His research interests are in understanding how developers use and comprehend different language features to improve language design.

\subsection*{Ali M. Keshk}

\includegraphics[width=1in,height=1.25in,trim={3.5cm 0 2.5cm 0},clip,keepaspectratio]{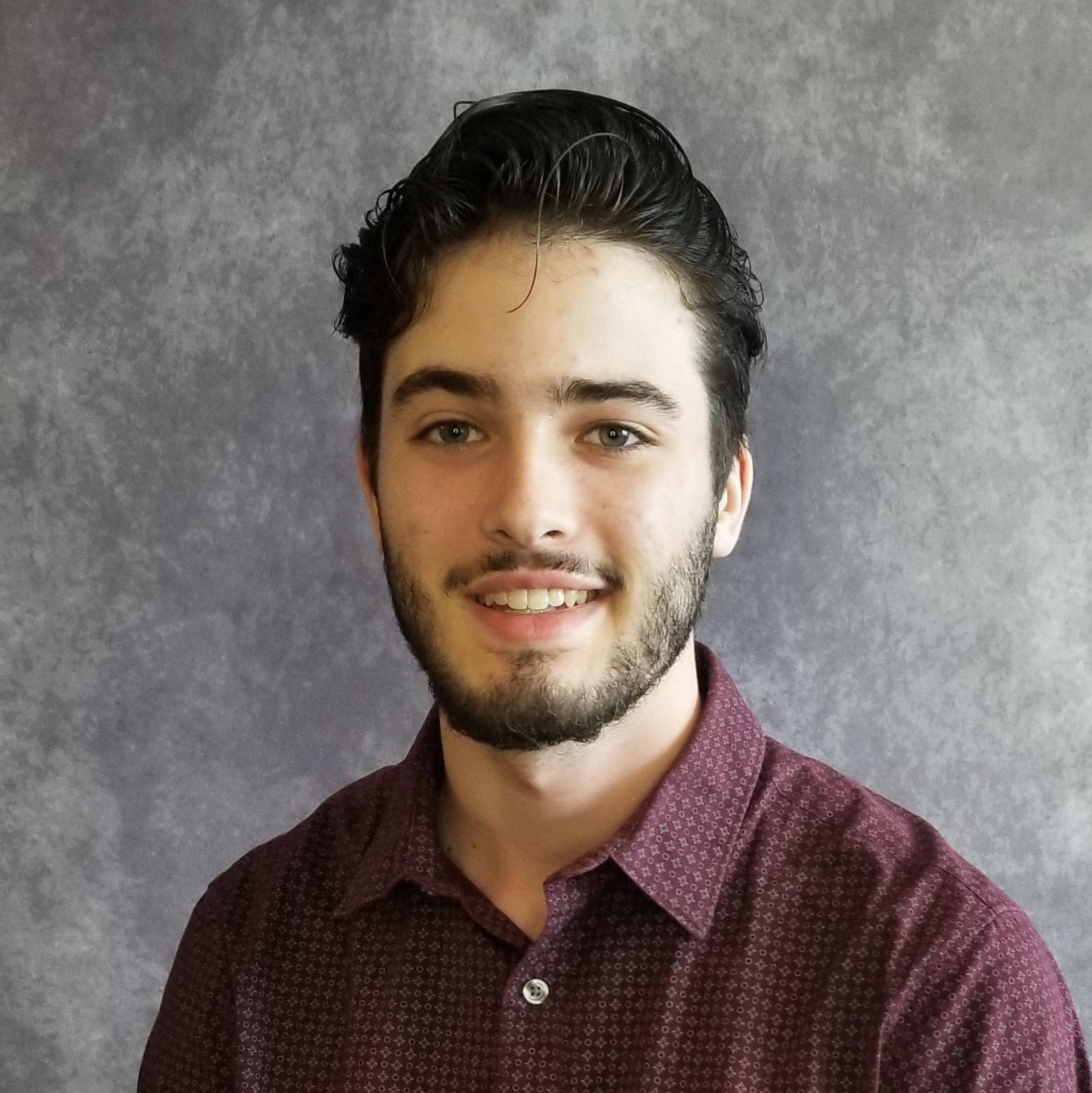}

Ali M. Keshk is an undergraduate student pursuing a Bachelor of Science in Computer Science at the University of Nebraska-Lincoln. With Dr. Robert Dyer as his advisor, his research primarily involved studying the use of method chaining across multiple programming languages.

\subsection*{Robert Dyer}

\includegraphics[width=1in,height=1.25in,clip,keepaspectratio]{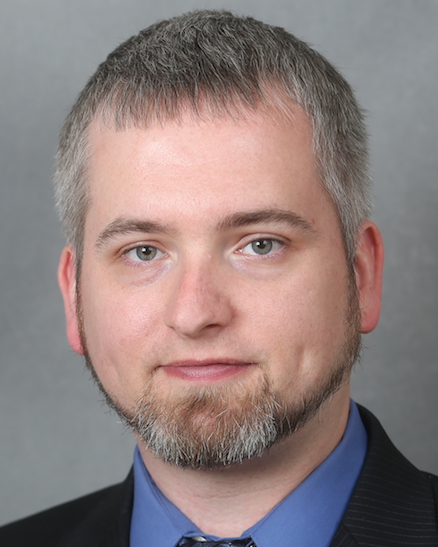}

Robert Dyer is an assistant professor at the University of Nebraska--Lincoln School of Computing. Previously he was an assistant professor at Bowling Green State University.  He received BS, MS, and Ph.D. degrees from Iowa State University.  His research is in applying mining software repository techniques to see how developers use programming languages and using data-driven approaches for language design: \url{https://go.unl.edu/rdyer}.

\subsection*{Hamid Bagheri}

\includegraphics[width=1in,height=1.25in,clip,keepaspectratio]{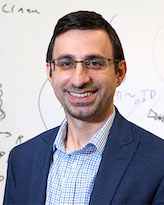}

Hamid Bagheri (Senior Member, IEEE) is an associate professor at the University of Nebraska--Lincoln School of Computing.  He received a Ph.D. degree in computer science from the University of Virginia. Prior to UNL, he was a postdoctoral researcher with the UC Irvine and MIT. His general research interests include the field of software engineering, and to date, his focus has spanned the areas of software analysis and testing, applied formal methods, and dependability analysis. He has received numerous awards for his research contributions, including the EPSCoR FIRST Award and the National Science Foundation CRII Award.

\end{document}